\documentstyle[aps,epsf,pra%
,twocolumn%
]{revtex}
\begin{document}
\draft
\title{On the driven Frenkel-Kontorova model: I. Uniform sliding
states and dynamical domains of different particle densities}
\author{Torsten Strunz and Franz-Josef Elmer}
\address{Institut f\"ur Physik, Universit\"at
   Basel, CH-4056 Basel, Switzerland}
\date{{\tt submitted to Phys. Rev. E, September, 1997, revised Jan. 98}}
\maketitle
\begin{abstract}
The dynamical behavior of a harmonic chain in a spatially periodic
potential (Frenkel-Kontorova model, discrete sine-Gordon equation) 
under the influence of an external force and a velocity proportional
damping is investigated. We do this at zero temperature for long
chains in a regime where inertia and damping as well as the
nearest-neighbor interaction and the potential are of the same order.
There are two types of regular sliding states: Uniform sliding
states, which are periodic solutions where all particles perform the
same motion shifted in time, and nonuniform sliding states, which
are quasi-periodic solutions where the system forms patterns of domains
of different uniform sliding states. We discuss the properties of
this kind of pattern formation and derive equations of motion for the
slowly varying average particle density and velocity. To observe
these dynamical domains we suggest experiments with a discrete ring
of at least fifty Josephson junctions.
\end{abstract}

\narrowtext

\section{Introduction}

60 years ago, Frenkel and Kontorova introduced a simple model which
has become popular in many fields of solid-state physics and
nonlinear dynamics \cite{kon38}. They invented their model in order
to describe the motion of a dislocation in a crystal \cite{INT:rem1}.
Meanwhile, the Frenkel-Kontorova (FK) model has become also a model
for an adsorbate layer on the surface of a crystal\cite{aub78,uhl88},
for ionic conductors \cite{pie81,aub83}, for glassy materials
\cite{pie81,uhl88,val88}, for charge-density-wave (CDW) transport
\cite{flo96}, for chains of coupled Josephson junctions
\cite{flo96,wat96}, and for sliding friction \cite{mcc89}.

The FK model is a chain of particles with mass $m$ coupled by a harmonic
nearest-neighbor interaction with stiffness $\kappa$. 
It is under the influence of an
external spatially periodic potential with periodicity $c$ and
strength $U_0$. Here we study the nonequilibrium behavior of 
the FK model driven by a force $\tilde{F}$. We assume
energy dissipation due to a usual damping force with a damping
constant $\eta$. After rescaling time and space,
one gets the following equation of motion in dimensionless units:
\begin{equation}
  \ddot x_j+\gamma\dot x_j=x_{j-1}+x_{j+1}-2x_j-b\sin x_j+F,
  \label{INT:eqm}
\end{equation}
where $\gamma\equiv \eta/\sqrt{\kappa m}$, $b\equiv
(2\pi/c)^2U_0/\kappa$, and $F\equiv (2\pi/c)\tilde{F}/\kappa$. The
time and space units are $\sqrt{m/\kappa}$ and $c/2\pi$, respectively.
We assume periodic boundary conditions, i.e.,
\begin{equation}
  x_{j+N}=x_j+2\pi M,
  \label{INT:bc}
\end{equation}
where $N$ is the number of particles and $M$ is an arbitrary integer.
Note that the periodic boundary condition implies that the average 
particle density $1/a$ is constant, i.e., 
\begin{equation}
  a=2\pi\,\frac{M}{N}.
  \label{INT:a}
\end{equation}
Due to symmetries $a$ can be restricted to $a\in [0,\pi]$ 
without loss of generality. 

The dynamical behavior of the FK model has already been studied in
several limits: (i) in the overdamped limit (i.e., $\ddot x_j$ can be
neglected) for large $N$ \cite{flo96,cop88}, (ii) in the limit of
zero damping and driving (i.e., $\gamma=F=0$) for $a/2\pi$ near
integer values allowing well separated $2\pi$ kinks \cite{pey84} as
well as for the most incommensurate value (i.e., the golden mean) of
$a/2\pi$ \cite{shi93}, and (iii) in the underdamped case for small
$N$ ($N\le 10$) \cite{wat96,ust93}. In a series of papers Braun {\em et al.}
have studied recently the underdamped dynamics of a generalized
FK model with $N>100$ but $a$ near zero and $\pi$ 
\cite{bra97a,pal97,bra97b}.

In this series of two papers, we study the underdamped FK model 
for large numbers
of particles (i.e., $N>100$). We do not restrict
ourselves to values of $a/2\pi$ near integer or half integer values
where the dynamical behavior can be described in terms of kinks.
The system has stationary, periodic, quasi-periodic, and chaotic
solutions. Of special interest is the transition from stationarity to
sliding, the so-called {\em pinning-depinning transition\/}, and the
backward process. In the first paper, we investigate the periodic and
quasi-periodic solutions. The second paper will be concerned with the
depinning-pinning transition between stationary states and
spatio-temporal chaos. Preliminary results have already been
published in a conference proceedings \cite{str96}.

In section~\ref{SD} we will see that the FK model spontaneously forms
spatial-temporal patterns as many other spatially extended systems
driven far from equilibrium \cite{cro93}. These patterns are caused
by bistability and instability of the {\em uniform sliding state\/}.
In the uniform sliding state all particles perform exactly the same
regular and periodic motion.  Different particles differ only in the
phase of this motion.  The phase difference of two neighboring
particles is everywhere the same.  That is, the density of particles
is on average constant along the chain.  The uniform sliding state is
the only nonstationary state in the overdamped limit \cite{mid92} but
it has also be studied in the underdamped case \cite{wat96,aub85a}. 

The discreteness of the chain leads to several {\em resonances\/} in
the underdamped limit \cite{wat96,ust93,aub85a,aub85b}. The consequence is
bistability. For spatially extended bistable systems it is well-known that
domain-like patterns are possible \cite{cro93}. In the FK model
these domains can be characterized by the average particle density
$1/a$ and the average particle velocity $v$. We find that states with
two or three different types of domains survive in the long-time
limit. The number and the width of the domains can vary, leading to a
quite large number of possible solutions. Because of different
average velocities in different domain types, the motion in the
domain solutions is quasi-periodic.  Assuming that $a$ and $v$ are
slowly varying functions in time and space, we derive an approximate
equation of motion for them. With its help we understand why the
domains do not disappear in the long-time limit and why there are not
more than three different domain types.  Furthermore it turns out
that the state in the long-time limit can be understood as a
spatially chaotic solution of a corresponding dynamical system. 
A special variant of these domain solutions is the {\em traffic-jam
state\/} where the particle velocity in one domain is zero
\cite{bra97a,pal97,bra97b}. 

The paper is organized as follows: In Section~\ref{VF} we 
derive two different
but mathematically equivalent formulas for the relation between the
force $F$ and the average sliding velocity $v$. 
In Section~\ref{US} the periodic solution (i.e., the
uniform sliding state) and its stability are discussed. The
domain-like states are investigated in Section~\ref{SD}. In
Section~\ref{CON} the main part of the paper concludes with some
remarks concerning possible experimental observations of these
domain-like states and similarities to other pattern forming
systems.  The Appendices describe our numerical and analytical 
scheme to obtain the uniform sliding state and to analyze its stability.

\section{\protect\label{VF}Average sliding velocity and effective
friction force}

The {\em average sliding velocity\/} $v$ of the chain reads
\begin{equation}
  v=\lim_{T\to\infty}\frac{1}{T}\int_0^T\frac{1}{N}
   \sum_{j=1}^N\dot x_j\,dt.
  \label{VF:v}
\end{equation}
Plotting the measured or calculated values of $v$ for different values
of the applied force $F$ one gets the so-called {\em velocity-force
characteristic\/}.  In CDW systems and Josephson-junction arrays it
corresponds to the current-voltage and voltage-current
characteristics, respectively. In the context of friction, $F$ can be
interpreted as the effective kinetic friction as a function of the
sliding velocity $v$.

There are two mathematically equivalent relationships between 
$F$ and $v$. The first
one can be obtained by taking the time average of the sum of
(\ref{INT:eqm}) over all particles:
\begin{equation}
  F=\gamma v+\lim_{T\to\infty}\frac{1}{T}\int_0^T\frac{b}{N}
   \sum_{j=1}^N\sin x_j(t)\,dt.
  \label{VF:Fv1}
\end{equation}
This formula assumes that the average acceleration of the chain is
zero. The other relationship can be derived from the fact that the
energy released during sliding has to be dissipated totally, i.e.,
\begin{displaymath}
 Fv=\gamma\lim_{T\to\infty}\frac{1}{T}\int_0^T\frac{1}{N}
   \sum_{j=1}^N\dot x_j^2\,dt.
\end{displaymath}
Using (\ref{VF:v}) we get:
\begin{equation}
 F=\gamma v\left[1+\lim_{T\to\infty}\frac{1}{T}\int_0^T
   \frac{1}{N}\sum_{j=1}^N\left(\frac{\dot x_j}{v}-1\right)^2dt
   \right].
  \label{VF:Fv2}
\end{equation}
This result clearly shows that in the case of no dissipation (i.e.,
$\gamma=0$) the chain may slide without any applied force. In
Section~\ref{INST} we will discuss the condition of this possibility.
Eq.~(\ref{VF:Fv2}) also shows that the mobility $v/F$ is always less
than or equal to $1/\gamma$, i.e., the mobility in the limit $b\to 0$.

\section{\protect\label{US}Uniform sliding}

Because of the symmetries of the equation of motion there exist
nonstationary solutions called {\em uniform sliding
states\/}. They are characterized by the fact that every
particle performs the same motion but shifted in time. That is,
$x_{j+1}(t)=x_j(t+T_1)$, for $j=1,\ldots,N$. Thus we need only one
function, the {\em dynamic hull function\/} $f$, to describe the
motion of all particles \cite{aub85a,cop88,flo96,wat96}:
\begin{equation}
  x_j(t)=\psi+aj+vt+f(\psi+aj+vt),
  \label{US:dhull}
\end{equation}
where $v$ is the average sliding velocity and $\psi$ is an arbitrary
phase. Because of the discrete translation symmetry of (\ref{INT:eqm})
the hull function is periodic, i.e.,
\begin{equation}
   f(\varphi+2\pi)=f(\varphi).
  \label{US:f1}
\end{equation}

Plugging the ansatz (\ref{US:dhull}) into the equation motion
(\ref{INT:eqm}) leads to a differential delay equation for the hull
function $f(\varphi)$:
\begin{eqnarray}
  v^2f''(\varphi)+\gamma v[1+f'(\varphi)]&=&f(\varphi+a)+f(\varphi-a)
   -2f(\varphi)\nonumber\\&&-b\sin[\varphi+f(\varphi)]+F.
  \label{US:DDEQ}
\end{eqnarray}
The average sliding velocity $v$ has to be chosen in such a way that
a $2\pi$-periodic function fulfills (\ref{US:DDEQ}). 
If $f(\varphi)$
is a solution then $f(\varphi+\psi)+\psi$ is also a solution. We make
the definition of the dynamic hull function unique by restricting the
solutions of (\ref{US:DDEQ}) to $2\pi$-periodic functions with zero
average, i.e.,
\begin{equation}
  \int_0^{2\pi}f(\varphi)\,d\varphi=0.
  \label{US:f2}
\end{equation}

For the uniform sliding state (\ref{US:DDEQ}), the relationships
(\ref{VF:Fv1}) and (\ref{VF:Fv2}) between $v$ and $F$ reduce to
\begin{equation}
  F=\gamma v+\frac{b}{2\pi}\int_0^{2\pi}\sin[\varphi
   +f(\varphi)]\,d\varphi
  \label{US:Fv1}
\end{equation}
and
\begin{equation}
  F=\gamma v\left(1+\frac{1}{2\pi}\int_0^{2\pi}[f'(\varphi)]^2
  d\varphi\right),
  \label{US:Fv2}
\end{equation}
respectively.

Instead of solving (\ref{US:DDEQ}) for a given value of $F$ it is
more convenient to replace $F$ by (\ref{US:Fv1}) or (\ref{US:Fv2})
and solve (\ref{US:DDEQ}) for a given value of $v$. From the solution
$f(\varphi)$ one obtains the corresponding $F$. In the overdamped
limit, it is well-known that $F$ is a monotonically increasing
function of $v$ \cite{mid92,flo96}. In the underdamped case {\em
resonances\/} lead to nonmonotonic velocity-force characteristics
\cite{US:rem1}.

\begin{figure}
\epsfxsize=80mm\epsffile{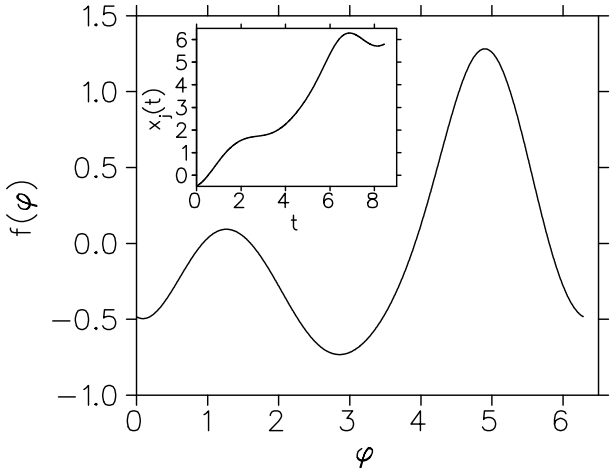}\vskip 2mm
\caption[Dynamic hull function]{\protect\label{f.hull} The
dynamic hull function $f(\varphi)$. The inset shows the corresponding
particle position as a function of time. The parameters are
$a/2\pi=(3-\sqrt{5})/2$, $b=2$, $v=0.75$, and $\gamma=0.5$. Because
the solution is obtained by the method described in
Appendix~\ref{HULL} the average sliding velocity $v$ is prescribed.
The corresponding value of $F$ is approximately $0.74$.}
\end{figure}

\begin{figure}
\epsfxsize=80mm\epsffile{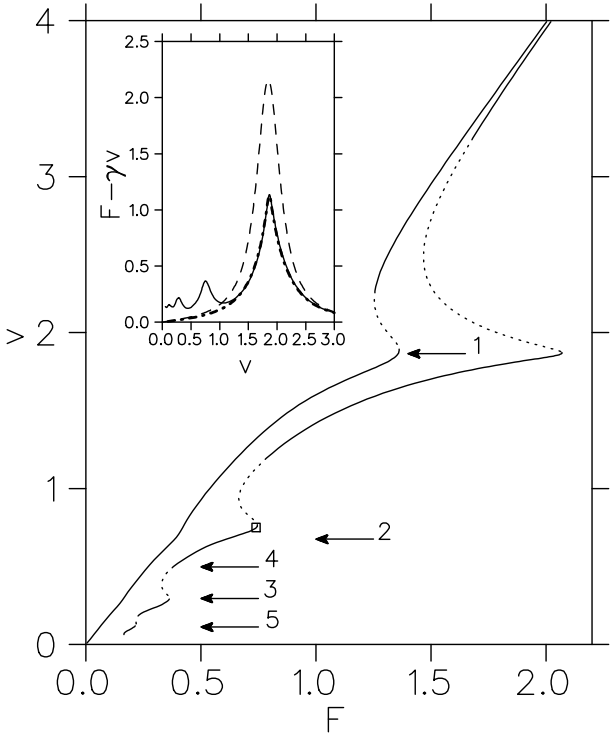}\vskip 2mm
\caption[Velocity-force characteristic]{\protect\label{f.vF}
Velocity-force characteristic of the uniform sliding state
(\ref{US:dhull}). Solid (dotted) lines indicate stable (unstable)
solutions. Dashed  and dashed-dotted lines in the inset denote
analytically obtained approximations given by (\ref{RES:F1}) and
(\ref{HULL:M1}), respectively.  The parameters are
$a/2\pi=(3-\sqrt{5})/2$, $\gamma=0.5$, and $b=1$ (left curve) and
$b=2$ (right curve and inset). The arrows denote the resonant values
of $v$ given by (\ref{RES:s.res}).  The numbers indicate the order
$n$.}
\end{figure}

\subsection{\protect\label{RES}Resonances}

We solve the hull-function equation (\ref{US:DDEQ}) numerically by
expanding $f$ into a Fourier series. The details are described in
Appendix~\ref{HULL}. Figure~\ref{f.hull} shows an example of a
dynamic hull function.

In the underdamped case the numerically obtained velocity-force
characteristics exhibits clearly peaks as can be seen in
figure~\ref{f.vF}. Near these peaks an increase of the driving force
$F$ by a considerable amount leads only to a slight increase of the
average velocity $v$.  In other words, the differential mobility,
$dv/dF$, is much lower than the mobility without any periodic
potential (i.e., $1/\gamma$).  The reason for this behavior is that
the additional energy is only partially turned into a larger kinetic
energy of the center of mass of the chain, whereas the main part is
turned into oscillatory motion of the particle due to resonances. 

In the frame of the center of mass which moves with average sliding
velocity $v$, the external potential leads to a time-periodic force
acting on the particle. The frequency of this force, the so-called
``washboard frequency'', is given by the velocity of the center of
mass divided by the period of the potential. Resonance occurs if the
washboard frequency is equal to the eigenfrequency of the phonon with
wave number $k=a$. To see this, we solve (\ref{US:DDEQ}) in the
approximation $\sin[\varphi+f(\varphi)]\approx\sin \varphi$. We get
\begin{equation}
  f(\varphi)\approx\frac{ib/2}{\omega^2(a)-v^2+i\gamma v}\,e^{i\varphi}
   +\mbox{c.c.},\label{RES:f1}
\end{equation}
where $\omega(k)$ is the phonon dispersion relation:
\begin{equation}
  \omega(k)=2\left|\sin\left(\frac{k}{2}\right)\right|.
  \label{RES:w}
\end{equation}
In order to get the corresponding value of $F$ one can use either 
(\ref{US:Fv1}) or (\ref{US:Fv2}). Although both equations are
equivalent, we get for the approximation (\ref{RES:f1}) different
results. Evaluating (\ref{US:Fv1}) leads to the obviously wrong
result $F=\gamma v$. This is can be understood from the fact that
(\ref{RES:f1}) is only the leading term of an expansion of the hull
function in powers of $b$. Thus $F=\gamma v+{\cal O}(b^2)$. Instead
of calculating the next order in $f$ one can use (\ref{US:Fv2}) which
leads to
\begin{equation}
  F=\gamma v\left(1+\frac{b^2/2}{[\omega^2(a)-v^2]^2+\gamma^2v^2}
   +{\cal O}(b^4)\right).
  \label{RES:F1}
\end{equation}
Note that $F$ has to be an even function of $b$ because the external
potential $b\cos x$ is an odd function of $x-\pi/2$. For large values
of $v$, (\ref{RES:f1}) approaches zero and therefore $F\to\gamma v$. 
That is, if the washboard slides very fast, the particles cannot
follow the fast pushing by the washboard. Thus the chain slides like
a rigid solid.  We call this state the {\em solid-sliding state\/}.

Figure~\ref{f.vF} shows that for large values of $b$ (or small values
of $\gamma$) this simple approximation (\ref{RES:F1}) overestimates
the strength of the resonance line. That is, there is an effective
damping, larger than $\gamma$, which increases with increasing
oscillation amplitude. This larger damping can be understood by
phonon coupling due to the nonlinearity in the equation of motion.
This coupling opens up additional channels for energy dissipation
yielding a higher effective damping constant. The inset of
Fig.~\ref{f.vF} shows that a remarkably good approximation of these
additional dissipation processes is given by the Galerkin
approximation $f(\varphi)=A\cos(\varphi+\psi)$. Projecting the
hull-function equation (\ref{US:DDEQ}) together with this ansatz onto
$\sin\varphi$ and $\cos\varphi$ leads to two transcendental equations
for $A$ and $\psi$. After elimination of $\psi$ one can parameterize
the velocity-force characteristic by the amplitude $A$ [see
Eq.~(\ref{HULL:M1}) in Appendix~\ref{HULL}].

In order to understand the other resonance peaks seen in
Fig.~\ref{f.vF} one has to keep in mind that the external potential
not only leads to a periodic driving force but also to a modulation
of the eigenfrequencies. Thus parametric processes make it possible
to excite other phonons. In the framework of a perturbation theory
with $b$ as the smallness parameter, the elementary processes
corresponding to these resonance lines are the decay of $n$
``washboard waves'' (wavenumber $a$, frequency $v$) into a single
phonon with wavenumber $k$ and frequency $\omega(k)$. Assuming
momentum and energy conservation one gets $k=na$ and $v=v_n$, where
$v_n$ is the {\em superharmonic resonance\/} frequency of order $n$:
\begin{equation}
  v_n\equiv \frac{\omega(na)}{n}.
  \label{RES:s.res}
\end{equation}

The positions of the first few resonances are shown in
Fig.~\ref{f.vF}. The agreement with the actual positions of the
resonance peaks is quite good. Near the superharmonic resonance of
order $n$ the $n$-th Fourier mode of the hull function has a maximum.
For example, the square in Fig.~\ref{f.vF} corresponds to the hull
function in Fig.~\ref{f.hull} which is clearly dominated by
$\exp(2i\varphi)$. Because of finite dissipation, superharmonic
resonances of higher order may be hidden behind a nearby resonance of
lesser order as, e.g., the fourth resonance in Fig.~\ref{f.vF}. 

The superharmonic resonance has been already investigated in the
literature, experimentally \cite{zan95} as well as theoretically
\cite{pey84,ust93,aub85a,aub85b}. The superharmonic resonance condition
(\ref{RES:s.res}) was observed first in numerical experiments by 
Aubry and de~Seze \cite{aub85a,aub85b}. They studied the FK model without
damping but with a very small driving force. They found that the
velocity of the center of mass did not increase linearly in time but
it was locked for finite time intervals at velocities given by
(\ref{RES:s.res}). They also studied at the first time the
underdamped and driven FK model \cite{aub85a}.
Peyrard and Kruskal found the same locking
phenomenon for the velocity of a $2\pi$-kink \cite{pey84}. In this
case the resonance frequencies are given by
\begin{equation}
  \tilde{v}_n\equiv \frac{\tilde{\omega}(na)}{n},\quad
   \mbox{with}\quad \tilde{\omega}(k)=\sqrt{b+\omega^2(k)}.
  \label{RES:s.res2}
\end{equation}
Ustinov {\em et al.\/} also observed resonance lines in numerical
simulations of the damped and driven FK model \cite{ust93}. Van der
Zant {\em et al.\/} have reported evidence of these resonances in
experiments with a ring of eight Josephson junctions \cite{zan95}. 

Ustinov {\em et al.\/} as well as van der Zant {\em et al.\/}
explained the found resonances by the following mechanism which leads
to (\ref{RES:s.res2}). The mechanism relies on the assumption that
$2\pi$-kinks are traveling in the ring. Most of the time the
particles are in a potential well. When a kink travels through a
particle it jumps into the neighboring well and oscillates. Because
of the periodic boundary conditions the jumps occur in equidistant
time steps. The distance between two kinks in terms of the number of
particles inbetween is given by $N/M=2\pi/a$. The kink velocity $c$
is related to the average sliding velocity $v$ by $v=ac$. 
Superharmonic resonances of order $n$ occur if the time interval
between two jumps, i.e., $(N/M)/c=2\pi/v$ is $n$ times the
oscillation time of the particles, i.e., $2\pi n/\tilde{\omega(k)}$,
where $\tilde{\omega(k)}$ given by (\ref{RES:s.res2}) is the
dispersion relation of the linearized equation of motion
(\ref{INT:eqm}). The wavenumber $k$ times the distance $N/M$ has to
be $2\pi n$, that is, $k=an$. According to this mechanism the
superharmonic resonance condition is therefore (\ref{RES:s.res2}).
This picture is valid only if the distance between two kinks, i.e.,
$N/M=2\pi/a$, is much larger than one. That is, the motion can be
described by kinks only if $a/2\pi$ is near an integer value. 
Thus, we expect that
(\ref{RES:s.res}) is valid for $a={\cal O}(1)$ and (\ref{RES:s.res2})
for $a\ll 1$. For example, the numerically obtained values of
positions of the resonance peaks reported by Watanabe {\em et
al.\/}~\cite{wat96} are more close to (\ref{RES:s.res}) than to
(\ref{RES:s.res2}) because $a>1$. 

\subsection{\protect\label{INST}Instabilities}

In order to discuss the instability of the uniform sliding state
(\ref{US:dhull}), one has to investigate the dynamics of small
perturbations $\delta x_j$. They are governed by the equation of
motion (\ref{INT:eqm}) linearized around (\ref{US:dhull}):
\begin{eqnarray}
 \delta\ddot x_j+\gamma\delta\dot x_j&=&\delta x_{j-1}+
   \delta x_{j+1}-2\delta x_j\nonumber\\&&-b\cos[aj+vt+f(aj+vt)]
   \delta x_j.
  \label{INST:leqm}
\end{eqnarray}
The periodic boundary condition (\ref{INT:bc}) turns into
$\delta x_{j+N}=\delta x_j$.
In accordance with the Floquet-Bloch theorem one can write any
solution of this equation as a sum of solutions of the form
\begin{equation}
  \delta x_j(t)=c_k(aj+vt)e^{\frac{2\pi i}{N}kj+\lambda_k t},\quad
   k=1,\ldots,N,
  \label{INST:f.sol}
\end{equation}
where $c_k(\varphi)$ is a $2\pi$-periodic function and $\lambda_k$ is
the so-called Floquet exponent. The uniform sliding state is stable
if the real part of $\lambda_k$ is negative for all values of $k$. 
There is always a solution with $\lambda_0=0$. It is the Goldstone
mode $\delta x_j=\partial_t x_j=v+vf'(aj+vt)\equiv c_0(aj+vt)$, which
follows directly from differentiating (\ref{US:DDEQ}).
Appendix~\ref{LINST} describes the scheme we have used to solve
(\ref{INST:leqm}) with the Floquet-Bloch ansatz (\ref{INST:f.sol})
numerically. The dotted lines in Fig.~\ref{f.vF} denote unstable parts
of the velocity-force characteristics.

Two different mechanisms may lead to an instability of the uniform
sliding state. The first one is {\em negative differential
mobility\/}, i.e., a negative slope in the velocity-force
characteristic. A small positive velocity fluctuation $v\to v+\delta
v$ accelerates the chain because the applied force is larger than the
force necessary to keep the new velocity $v+\delta v$ constant.

The second type of instabilities is caused by {\em parametric 
resonance\/}. In the framework of multi-phonon process it corresponds
to the decay of $n$ washboard waves into two phonons with wavenumber
$na/2\pm q$. Parametric resonance of order $n$ can be
expected for values of the average sliding velocity $v$ which are
given by
\begin{equation}
  v_n^P(q)=\frac{\omega(na/2+q)+\omega(na/2-q)}{n}.
  \label{INST:p.res}
\end{equation}
Because parametric resonance is a threshold phenomenon, the amplitude
$b$ of the washboard wave has to exceed a critical value which is
proportional to $\gamma^{1/n}$ \cite{cro93,wei97}. This is true only for
values of $v$ between the minimum and the maximum of
(\ref{INST:p.res}). For velocities outside this interval parametric resonance is
still possible but the threshold increases.
For zero damping the uniform sliding state is unstable for 
any value of $v$ below a certain critical value $v_c$ which is
approximately calculated in Appendix~\ref{LINST}:
\begin{equation}
 v_c\approx\sqrt{16\cos^2\left(\frac{a}{4}\right)+2b}.
  \label{INST:vc}
\end{equation}
The actual value of $v_c$ obtained from the numerical stability analysis 
agrees very well with this formula even for large values of $b$.
The numerical value of $v_c$ is less than (\ref{INST:vc}) but deviates not 
more than 10 percent for $b<4$.

\begin{figure}
\epsfxsize=80mm\epsffile{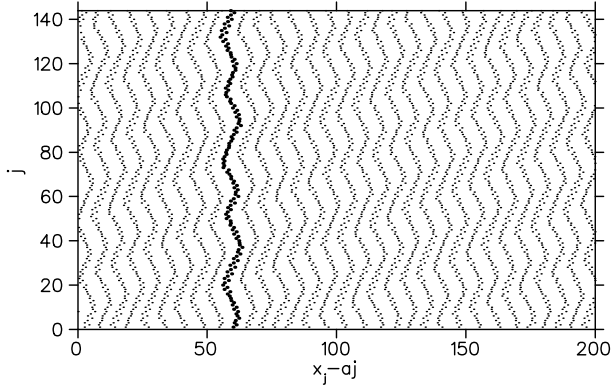}\vskip 2mm
\caption[Dynamical domains]{\protect\label{f.mdom} Dynamical domains
of different particle densities. A series of snapshots taken at
equidistant time steps ($\delta t=2\pi/v$) are shown. Each dot
denotes the position of a particle. A particular zig-zag shaped
snapshot is highlighted. The zigs and zags correspond to two
different kind of domains that are characterized by uniform
sliding.  The parameters are $N=144$, $M=55$, $b=2$, $\gamma=0.5$,
and $F=0.8$.}
\end{figure}

\section{\protect\label{SD}Sliding domains}

Near resonance peaks the system is bistable and has therefore the
opportunity to organizes itself into {\em domains\/} of different
uniform sliding states. Figure~\ref{f.mdom} shows a typical example
with ten domains. There are only two types of domains. Each domain
is characterized by uniform sliding. That is, in each domain the
particle motion is given by (\ref{US:dhull}) but the hull function
$f$ and the value of $a$ and $v$ are different in each domain.  One
can say that the domains are characterized by different particle
densities $1/a$. Neighboring domains are separated by domain walls
(fronts) of finite size.  Conservation of the number of particles
implies that a front has to move with the velocity
\begin{equation}
  v_{\rm front}=\frac{a_2v_1-a_1v_2}{a_2-a_1},
  \label{SD:vDW}
\end{equation}
where the average particle distance and the average particle
velocity of each domain type is given by $a_{1,2}$ and $v_{1,2}$,
respectively. From the viewpoint of the particles, we can express
the front velocity in terms of how fast the front travels from one
particle to the next. It is given by
\begin{equation}
  c=\frac{v_1-v_2}{a_2-a_1}.
  \label{SD:c}
\end{equation}
Because (\ref{SD:vDW}) and (\ref{SD:c}) are symmetric in the
exchange of the indices, all fronts propagate with the same
velocity leaving the widths of the domains constant. The numbers
$N_{1,2}$ of particles in each type of domain fulfill the following
constraints:
\begin{equation}
  N_1+N_2=N\quad\mbox{and}\quad a_1N_1+a_2N_2=2\pi M=aN
  \label{SD:N12}
\end{equation}
Because of $0<N_{1,2}<N$ the particle density for one type of the
domains is larger than $1/a$ whereas for the other type it is
less than $1/a$. In the following the type with the larger
density will be number one. Thus,
\begin{equation}
  a_1<a<a_2.
  \label{SD:a}
\end{equation}
The average sliding velocity $v$ of a domain state is given by
\begin{equation}
  v=\frac{v_1N_1+v_2N_2}{N}.
  \label{SD:v}
\end{equation}

A domain-type state is in general a quasiperiodic motion with three
frequencies: $v_1$ and $v_2$ from the periodic motion in each domain
type and $2\pi c/N$ from the cyclic motion of the fronts through the
system.

\begin{figure}
\epsfxsize=80mm\epsffile{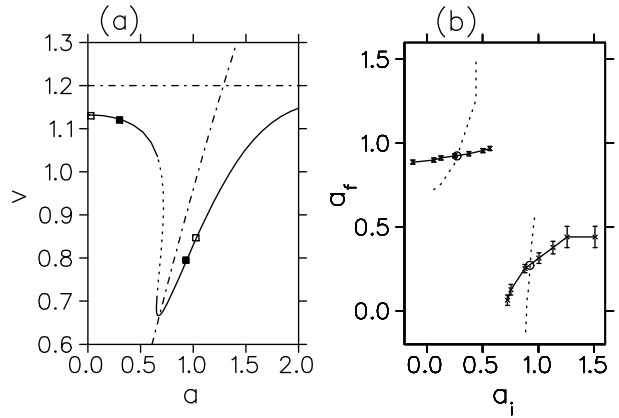}\vskip 2mm
\caption[Velocity-distance characteristic]{\protect\label{f.va} (a)
The velocity-distance characteristic for the uniform sliding states.
Solid (dotted) lines indicate stable (unstable) states. The
dash-dotted horizontal (tilted) line denotes $v=F/\gamma$ (the main
resonance $v_1$). The open (filled) squares denote the initial
(final) domain states of the simulation shown in
Fig.~\ref{f.td.relax}.  (b) The numerically obtained values of the
average particle distances $a_f$ behind the front for a given value 
$a_i$ in front of the front. Those data points are connected by a
solid line where a continuous function is expected. The dotted line
denotes the inverse function. The circle near the intersection of
both functions denote the values of $a_1$ and $a_2$ of the
numerically found two-domain solution.  The parameters are $b=0.5$,
$\gamma=0.5$, and $F=0.6$.}
\end{figure}

\subsection{\protect\label{TDS}The two-domain state}

To understand this kind of pattern formation we investigate in detail
two-domain states. First, we have to discuss the relationship between
the average sliding velocity $v$ and the average particle distance
$a$ for the uniform sliding state at a fixed value of $F$, i.e., the
{\em velocity-distance characteristic\/}. A typical example is shown
in figure~\ref{f.va}(a). As in the velocity-force characteristic,
resonances are responsible for folds.

\begin{figure}
\epsfxsize=80mm\epsffile{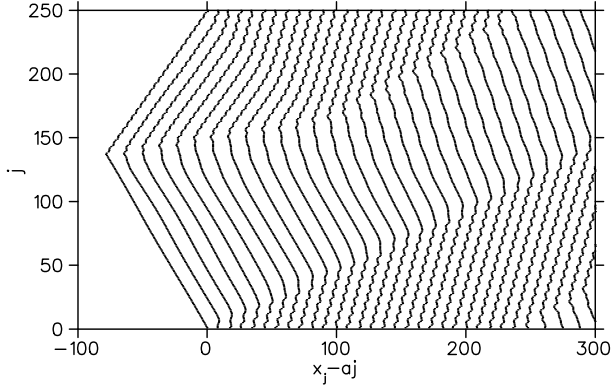}\vskip 2mm
\caption[Dynamics of a two-domain state]{\protect\label{f.td.relax}
Snapshots of the evolution of a two-domain state taken at equidistant
time steps ($\Delta t=12$). The presentation is the same as in
Fig.~\ref{f.mdom}. The initial state is a two-domain state with
$a_1=2\pi/100$ and $a_2=2\pi/5$. The initial positions and velocities
of the particles are calculated by using (\ref{US:dhull}).  The
parameters are $N=250$, $M=25$, $b=0.5$, $\gamma=0.5$, and $F=0.6$. 
}
\end{figure}

A two-domain state is completely characterized by two points on the
stable branches of the velocity-distance characteristic which
fulfill (\ref{SD:a}). The velocity $c$ of the front is the
slope of the line connecting both points (see Fig.~\ref{f.cg}). The
sizes of the domains are determined by the solutions of
(\ref{SD:N12}).  

From this consideration one would expect a continuous family of
two-domain states parameterized by two real numbers. But this is not
the case, as figure~\ref{f.td.relax} clearly shows. In our numerical
simulations we always found that the system selects dynamically the
same pair of points on the velocity-distance characteristic. This is
true even if $a$ is changed as long as $a\in(a_1,a_2)$. How does the
system selects a certain pair of values for $a_{1,2}$? A careful
inspection of Fig.~\ref{f.td.relax} reveals that behind the fronts
new domain states are selected. These states are independent of the
initial states. The interface between the new domain state and the
old one does not form a front. It smears out and in the long-time
limit the domain states approach to uniform densities with
well-defined values of $a$. Numerical experiments show that the state
$a_f$ behind the front is uniquely defined by the state $a_i$ in
front of the front [see Fig.~\ref{f.va}(b)]. Hence there is
functional relation between them:
\begin{equation}
  a_f=A(a_i).
  \label{TDS:A}
\end{equation}
Together with (\ref{SD:c}) we have a uniquely defined relation
between $c$, $a_i$, and $a_f$. This behavior fits into the general
picture of front propagation in bistable systems \cite{cro93} (see
also Sec.~\ref{QCD}). If we know the function $A$ we can calculate
the values of $a_1$ and $a_2$ by solving $A^{-1}(a)=A(a)$ [see also
Fig.~\ref{f.va}(b)]. Note that the values of $a_1$ and $a_2$ are in
general irrational \cite{rem2}. The value of $a$ determines only the
sizes of the domains. Whether the chain is commensurate or
incommensurate is irrelevant. But numerically we have never found
such states for values of $a$ near integer multiples of $2\pi$.

\begin{figure}
\epsfxsize=80mm\epsffile{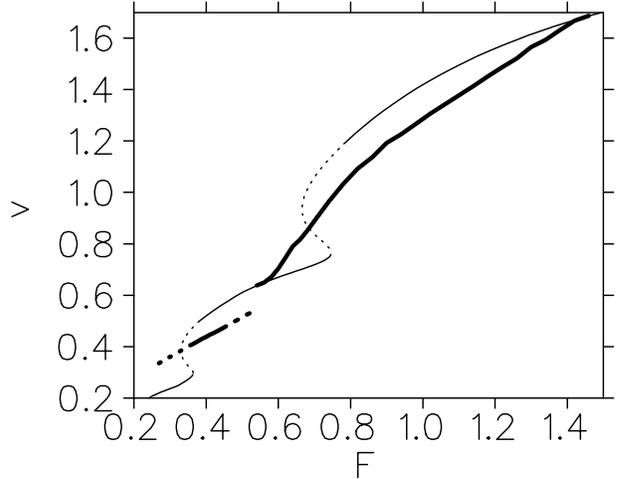}\vskip 2mm
\caption[Velocity-force characteristic of a two-domain state]
{\protect\label{f.vF.td} Velocity-force characteristics for the
uniform sliding states and the two-domain states. Thick (thin) lines
denote two-domain (uniform sliding) states. Solid (dotted) lines
indicate stable (unstable) solutions. The parameters are $N=2584$,
$M=987$, $b=2$, and $\gamma=0.5$. }
\end{figure}

The selected values of $a_1$ and $a_2$ are of course functions of the
applied force $F$. Figure~\ref{f.vF.td} shows the velocity-force
characteristic of the two-domain states. By varying $F$ they
disappear because of two reasons. First, one of the domains shrinks
to zero and the velocity-force characteristics of the two-domain
state and uniform sliding state come together. Secondly, $a_1$ or
$a_2$ may move onto an unstable branch of the velocity-distance
characteristic.  Thus, the two-domain state still exists but it has
become unstable.  In Fig.~\ref{f.vF.td} the velocity-force
characteristic of this unstable two-domain state is denoted
schematically by dotted extensions.

\subsection{\protect\label{QCD}Quasi-continuum description of the
front}

We have seen that the state of the domains can be
described by (\ref{US:dhull}). They are characterized by $a_{1,2}$
and $v_{1,2}$. To describe the fronts in the same way we
assume that $a$ and $v$ are continuous functions which are varying
slowly in space and time. The space coordinate is the particle index
$j$. It becomes a real variable in a quasi-continuum description.
Thus, we write $x_j(t)=x(j,t)$. The discrete Laplacian
$x_{j+1}+x_{j-1}-2x_j$ can be written as an infinite series of
differential operators, i.e.,
\begin{equation}
  x_{j+1}(t)+x_{j-1}(t)-2x_j(t)=4\sinh^2\left(\frac{1}{2}\partial_j
   \right)x(j,t).
  \label{QCD:d2}
\end{equation}
We generalize the ansatz (\ref{US:dhull}) by assuming that $\varphi$ is
a function of $j$ and $t$, i.e.,
\begin{equation}
  x(j,t)=\varphi(j,t)+f\bigl(\varphi(j,t)\bigr),
 \quad f(\varphi+2\pi)=f(\varphi),
  \label{QCD:ansatz}
\end{equation}
where $f$ is the solution of the hull function equation
(\ref{US:DDEQ}). Now we define local values of $a$ and $v$ by
\begin{equation}
  a\equiv \partial_j\varphi,\quad\text{and}\quad v\equiv\partial_t\varphi.
  \label{QCD:av}
\end{equation}
Plugging the ansatz (\ref{QCD:ansatz}) into the equation of motion
(\ref{INT:eqm}) and averaging over the phase $\varphi$ we get
\begin{mathletters}\label{QCD:eqm}
\begin{eqnarray}
  \partial_tv&=&D(\partial_j)\partial_ja+F-F_U(a,v),\label{QCD:eqm.v}\\
  \partial_ta&=&\partial_jv\label{QCD:eqm.a},
\end{eqnarray}
where
\begin{equation}
  D(x)\equiv\left(\frac{\sinh x/2}{x/2}\right)^2=1+\frac{x^2}{12}
   +{\cal O}(x^4),
  \label{QCD:D}
\end{equation}
\end{mathletters}
and $F_U(a,v)$ is the velocity-force characteristic of the uniform
sliding state given by (\ref{US:Fv1}). 
Eq.~(\ref{QCD:eqm.v}) is only approximately correct because we have
assumed that the hull function $f$ does not depend on $a$ and $v$.
Furthermore, the ansatz (\ref{QCD:ansatz}) cannot be exact in a
front. Nevertheless, the approximation (\ref{QCD:eqm}) is correct in
leading order of a multiple-scale perturbation theory
\cite{whi74,zwi89}.

\begin{figure}
\epsfxsize=80mm\epsffile{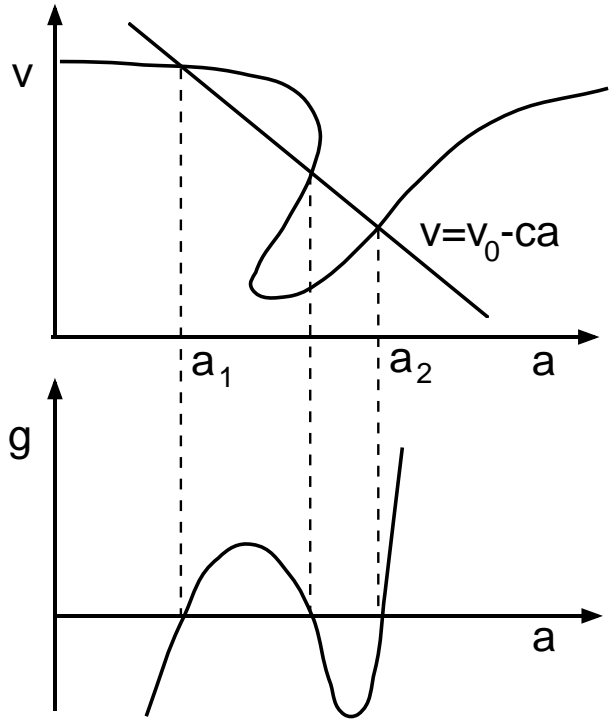}\vskip 2mm
\caption[Driving force]{\protect\label{f.cg}
Schematical drawing of the velocity-distance characteristic and the
corresponding nonlinearity $g$.
}
\end{figure}

In general eq.~(\ref{QCD:eqm}) cannot be solved analytically.
But we are able to discuss the front solutions qualitatively. 
Assuming stationarity of the front in the co-moving frame 
we get 
\begin{equation}
  a(j,t)=a(j-ct),\quad v(j,t)=v(j-ct),
  \label{QCD:wall}
\end{equation}
where $c$ is the front velocity (\ref{SD:c}). From
(\ref{QCD:eqm.a}) we get $-ca'=v'$ which can be integrated leading to
\begin{equation}
  v=v_0-ca.
  \label{QCD:wall.v}
\end{equation}
Plugging (\ref{QCD:wall}) and (\ref{QCD:wall.v}) into (\ref{QCD:eqm.v})
and keeping only the first two terms of $D$ yields
\begin{equation}
  (1-c^2)a'+\frac{1}{12}a'''+g(a,v_0,c)=0,
  \label{QCD:wall.a}
\end{equation}
with
\begin{equation}
   g(a,v_0,c)\equiv F-F_U(a,v_0-ca).
  \label{QCD:g}
\end{equation}

Eq.~(\ref{QCD:wall.v}) means a straight line in the velocity-distance
characteristics (see Fig.~\ref{f.cg}). Front solutions exist for
those values of $v_0$ and $c$ for which (\ref{QCD:wall.v}) intersects
the velocity-distance characteristic of the uniform sliding state 
three times. The two outside intersection points have to lead to 
stable uniform sliding states (characterized by $a_1$ and $a_2$)
whereas the inner point has to belong to an unstable uniform 
sliding state. This is the reason
why two-domain and multi-domain solutions appear only near resonance
points where the velocity-force characteristic has a negative slope
(see Fig.~\ref{f.vF.td}). Because there is no resonance for $a/2\pi$
integer, we understand why we have not found two-domain solutions
for values of $a$ close to integer multiples of $2\pi$.

If the requirements on $v_0$ and $c$ are fulfilled, the nonlinear term $g$ in
(\ref{QCD:wall.a}) will have three nodes and will be N-shaped (see
Fig.~\ref{f.cg}). A front solution is a heteroclinic orbit of
(\ref{QCD:wall.a}) which goes from $a_1$ to $a_2$ or vice versa.
Thus, we are looking for solutions with boundary conditions
$a(-\infty)=a_{1,2}$ and $a(\infty)=a_{2,1}$. A heteroclinic orbit
occurs only if the unstable manifold of the fixed point $a_{1,2}$ is
the stable manifold of $a_{2,1}$. This is possible only on a
one-dimensional manifold in the parameter space of $v_0$ and $c$.
Thus (\ref{TDS:A}) is justified. 

To calculate the stable and unstable manifolds we
linearize (\ref{QCD:wall.a}) around the fixed points $a_{1,2}$. For
the perturbation $\delta a\equiv a-a_{1,2}$ we make the ansatz $\delta
a=\exp(\lambda j)$ which leads to the characteristic polynomial
\begin{equation}
  \frac{1}{12}\lambda^3+(1-c^2)\lambda+\partial_ag(a_{1,2})=0.
  \label{QCD:wall.a.lin}
\end{equation}
Because of $\partial_ag(a_{1,2})>0$ (see Fig.~\ref{f.cg}) there is
one negative root $\lambda_1<0$. If $(1-c^2)^3+(3\partial_ag/4)^2>0$
the two other solutions are conjugated complex with a real part that
is just $-\lambda_1/2$. Numerically we always found subsonic front
motion (i.e., $|c|<1$) leading to an unstable manifold that spirals
out of the fixed point. Thus, the precursor of the front is
non-oscillatory whereas its tail is oscillatory because the particles
have inertia and therefore respond with an exponentially decreasing
oscillation after an acceleration or deacceleration. 

In order to verify this qualitative picture numerically one has to
extract $a$ and $v$ from the data. In principle this could be done by
local fits of the dynamic hull function from which we obtain
$\varphi(j,t)$ and subsequently $a$ and $v$. But this is a very
tedious way which is not necessary because we are only interested in
the qualitative form of the shape of the front. The following simple
method is sufficient for this task. For uniform sliding states it
leads to values of $a$ and $v$ which are identical with the exact
ones.  We introduce for each particle a sequence of times $t_{n,j}$
defined by $x_j(t_{n,j})=(2n+1)\pi$ and $\dot x_j(t_{n,j})>0$. From
these sequences we get the following approximations of $v$ and $a$:
\begin{equation}
  v(j,t)\approx\frac{2\pi}{t_{n,j}-t_{n-1,j}},\quad
  a(j,t)\approx v(j,t)(t_{m,j-1}-t_{n,j}),
  \label{QCD:va.exp}
\end{equation}
where $n$ and $m$ are chosen in such a way that (i)
$t_{n-1,j}<t<t_{n,j}$ and (ii) $t_{m,j-1}$ is the time closest to
$t_{n,j}$, i.e., $|t_{m,j-1}-t_{n,j}|= \min_{\tilde{n}} 
|t_{\tilde{n},j-1}-t_{n,j}|$. This definition of $a$ is
necessary in order to avoid spurious values which differ from the
expected value by $\pm 2\pi$. 

\begin{figure}
\epsfxsize=80mm\epsffile{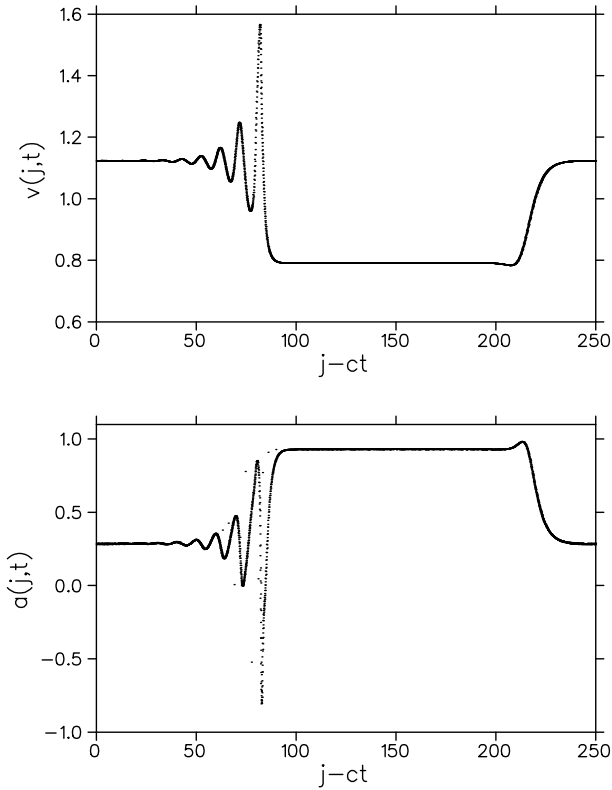}\vskip 2mm
\caption[Quasi-continuum description]{\protect\label{f.qcd}
Quasi-continuum description of the two-domain solution. The values of
$v$ and $a$ are obtained from the formulas (\ref{QCD:va.exp}).  20
snapshots taken at different times and shifted by $ct$ are superposed
in order to get the details of the fronts. The value of $c$ is chosen
in such a way that the superposition yields a smooth curve.  The best
value of $c$ is $0.516$. The parameters are the same as in
Fig~\ref{f.td.relax}.  }
\end{figure}

For a fixed value of $t$ one gets a snapshot of $v$ and $a$. The
superposition of many snapshots shifted by $ct$ gives the impression
of a smooth curve (see Fig.~\ref{f.qcd}). We have tuned $c$ until the
curves are as smooth as possible. It turned out that this method is a
very accurate way to measure the front velocity $c$. The
precursors and tails of both fronts are non-oscillating and
oscillating, respectively. Furthermore, the oscillatory tail of the
front on the left decays roughly two times slower than the precursor
of the front on the right. Both observations are in full agreement with
our analytical reasoning above.

\subsection{\protect\label{MDS}Multi-domain states}

\begin{figure}
\epsfxsize=80mm\epsffile{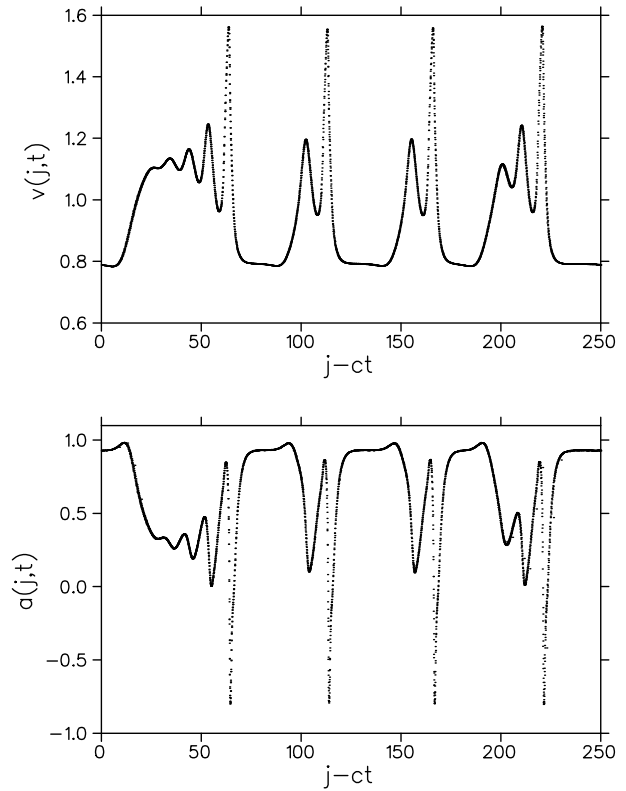}\vskip 2mm
\caption[Multi-domain solution]{\protect\label{f.mds}
An example of a multi-domain state. The parameters are the same as in
Fig~\ref{f.td.relax}.}
\end{figure}

Starting from an arbitrary initial condition one gets either a
uniform sliding state (if a stable one exists) or a multi-domain
state but only rarely a two-domain state. This is especially true for
large systems. It is a very general behavior of bistable spatially
extended systems, at least for the initial phase of the dynamics. 
Different parts of the system establish themselves independently into
one of the bistable states. It is therefore natural that several
domains occur.  After the initial formation of a multi-domain state,
domains may shrink and eventually disappear on a slow time scale.
This can be understood by the fact that an attractive force between
the fronts exist \cite{kaw82}. This force is caused by the overlap of
the precursor and the tail of the neighboring fronts. It can be
calculated by singular perturbation theory \cite{cro93,kaw82}. In the
case of non-oscillating precursors and tails the force decreases
exponentially with distance \cite{kaw82}. In our case where the tail
is oscillating, the resulting force is also oscillating
\cite{cro93,elp88}. Therefore equilibrium positions are possible
where a pair of two fronts form a bound molecule.  The available
distances between the fronts are quantized. Figure~\ref{f.mds} shows
an example where this quantization is clearly seen. In accordance
with Shilnikov's theorem such molecules are possible only if the
oscillatory tail decays slower than the nonoscillatory one
\cite{arn81,arn82} which is indeed the case (see Sec.~\ref{QCD}). As
a consequence the system shows spatial chaos.

For values of $F$ near higher order resonances the velocity-distance
characteristic of the uniform sliding state shows several resonance
folds. Thus, multi-domain states are possible which are built up from
more than two different domain types. All our numerical experiments
have shown that no matter how many domain types occur in the
transient, at the end (i.e., in the long-time limit) only two or
three domain types survive. Furthermore, all fronts travel with the
same velocity $c$.  There is a simple argument why more than three
domain types are inconsistent with the last fact. Consider the case 
of four different domain types with $a_1<a_2<a_3<a_4$ and
$v_1>v_2>v_3>v_4$. Between two consecutive values $a_i$ and $a_{i+1}$
there should be no additional stable state and only one unstable
state. Thus, if two of such domain types are neighbors the states are
functionally related in accordance with (\ref{TDS:A}). Let us assume
a sequence of domains from right (domain type 1) to left (domain type
4). Using (\ref{TDS:A}) we get $a_{i+1}=A(a_i)$. This sequence is
therefore uniquely determined by $a_1$. Now the condition that all
fronts have the same speed cannot be fulfilled because there is only
one free variable but (at least) two conditions, namely,
$c_{12}=c_{23}=c_{34}$. Thus, four different domain types are in
contradiction with the observation concerning the front velocities. 

For $F<b$ domain-like solutions occur where in one domain
state the particles sit in potential wells (average particle distance
$a$ is an integer multiple of $2\pi$, e.g., $a=0$) and do not move.
Such so-called traffic-jam solutions have been already reported by
Braun {\em et al.\/} \cite{bra97a,pal97,bra97b}. They mainly occur
near integer values of $a/2\pi$ where the domain-like states discussed
in this section are not possible.

\section{\protect\label{CON}Conclusion}

In this paper the regular sliding states (periodic and quasi-periodic
solutions) of the FK model are investigated for large chains
($N>100$) and for values of $\gamma$, $b$, and $F$ where all forces
in the equation of motion (\ref{INT:eqm}) are of the same order. 

Instead of classifying these attractors to be either periodic or
quasiperiodic, a more informative distinction is whether their
locally average particle density is uniform or not. In a periodic
state the density is uniform because all particles perform the same
motion only shifted in time.  The motion of the whole chain is
completely determined by a single periodic function, the dynamical
hull function. The velocity-force characteristics of the periodic
attractors show peaks at certain values of the velocity. These peaks
are caused by resonances of the ``washboard'' wave. In the frame of
the center of mass of the chain the external potential can be seen as
a wave (washboard wave) with wave number $a$ (which is the inverse
average particle density) and frequency $v$ (which is the average
sliding velocity). Resonances occur for those values of $v$ where $n$
washboard waves are able to decay into a phonon [wavenumber $k$,
frequency $\omega(k)$] in accordance with ``momentum'' and ``energy''
conservation [i.e., $k=na$ and $\omega(k)=nv$]. A similar process
with a decay into two phonons (parametric resonance) explains why the
uniform sliding state may be unstable even though the differential
mobility is positive.

Quasiperiodic states emerge from instabilities of uniform sliding
states. They are characterized by domains of different average
particle densities. In each domain the average particle velocity is 
uniform and {\em nonzero\/}. The walls between neighboring domains
move all with the same velocity. This domain-wall or front velocity
is different from the velocity of the center of mass. Thus, each
particle goes through all domains. In a quasiperiodic attractor not
more than three different domain types are possible. For chains close
to the most commensurate case (i.e., $a$ is an integer multiple of
$2\pi$) the average particle velocity is zero in one domain. In this
so-called traffic-jam state \cite{bra97a} the particles are
alternately switch between stationarity and sliding. Because the
traffic-jam state exists also for finite temperatures 
\cite{bra97a,bra97b} states with different sliding domains will
presumably survive finite temperatures.

In this paper we have developed a quasi-continuum description of the
FK model on the basis of slowly varying locally averaged inverse
density $a$ and velocity $v$. That is, $a$ and $v$ slowly depend on
the particle index $j$ and the time $t$. A set of partial
differential equations [Eqs.~(\ref{QCD:eqm})] governs the dynamics of
the variables (assuming $j$ to be a continuous space variable). It is
not possible to solve these partial differential equations
analytically. But it is quite helpful to understand (i) why the FK
model organizes itself into domain-like states with domain states
which are not determined from outside, (ii) why there are not more
than three different domain types, (iii) why the size of the domains
is quantized, (iv) why states with several domains of the same type
are possible, (v) why a multi-domain state can be seen as an example
of spatial chaos, and (vi) why for a fixed value of the external
force $F$ the number of stable states increases exponentially with
$N$. The huge number of stable states leads to multi-hysteretic
behavior like in a ferromagnet where the position in the
velocity-force characteristic strongly depends on the history.

The phenomenon of domain-like states where each domain is
characterized by a spatially periodic but stationary solution has
already been found in hydrodynamical pattern formation
(Taylor-Couette system \cite{bax86} and Rayleigh-B\'enard system
\cite{heg92}). Theoretically this behavior has been modeled by 
nonlinear phase equations \cite{cro93,bra89,rie90}. In our case the
domain states are periodic in space and time, they are traveling
waves. 

In experiments where locally resolved measurements
are not possible,  the most important consequence of the domain-like 
sliding states are (i) quasiperiodicity in the time signals like the 
velocity of the center of mass, (ii) flattening of the resonance peaks
(see Fig.~\ref{f.vF.td}), and (iii) multihysteretic behavior.
Many of these features should disappear for small values of $N$
if they are caused by domain-like states.
Because the inertia term is important for these states, we do not
expect it in CDW-systems. In adsorbate layers and ionic conductors
the appearance of them should be possible but it may be difficult to
drive them {\em uniformly\/} and to measure the
velocity-force characteristic. The ideal system to check our theory
are Josephson-junction arrays because the force and the velocity
correspond to the driving current and the voltage, respectively.
Furthermore, the damping constant $\gamma$ and the number of
junctions $N$ can be chosen by the fabrication process. 
The average distance $a=2\pi M/N$ is also easily accessible because 
the number $M$ of flux quanta can be chosen by the initial preparation
of the system.
Thus, for rings of more than 50 Josephson junctions we predict
the occurrence of domain-like sliding states.  

\acknowledgements

We thank H. Thomas for critical reading of the manuscript. This work
was supported by the Swiss National Science Foundation.

\appendix

\section{\protect\label{HULL}Numerical and analytical approximations
of the hull-function equation (\ref{US:DDEQ})}

In this appendix we describe the numerical scheme we have used to
solve the differential delay equation (\ref{US:DDEQ}) for the dynamic
hull function $f(\varphi)$. In the simplest case this scheme leads also
to a nonlinear analytical approximation. 

Because of (\ref{US:f1}) and (\ref{US:f2}) we expand the dynamic hull
function $f$ into a Fourier series
\begin{equation}
  f(\varphi)=\sum_{m=1}^\infty f_me^{im\varphi}+\mbox{c.c.}
  \label{HULL:Fs}
\end{equation}
In the numerical approximation we replace this infinite series by a
finite one with a cutoff $M$, i.e., $\sum_{m=1}^\infty\to
\sum_{m=1}^M$. Plugging this ansatz into the differential delay
equation (\ref{US:DDEQ}) leads to a set of nonlinear algebraic
equations for the coefficients $\{f_1,f_2,\ldots,f_m,\ldots\}$:
\begin{equation}
  [(mv)^2-\omega^2(ma)-i\gamma mv]f_m=bI_m(f_1,f_2,\ldots,f_j,\ldots),
  \label{HULL:eq}
\end{equation}
where $\omega(k)$ is the phonon dispersion relation (\ref{RES:w}) and 
\begin{equation}
  I_m(f_1,f_2,\ldots,f_j,\ldots)\equiv \frac{1}{2\pi}\int_0^{2\pi}
   \sin[\varphi+f(\varphi)]e^{-im\varphi}d\varphi.
  \label{HULL:I}
\end{equation}
The driving force $F$ does not appear in the set of
equations (\ref{HULL:eq}). That is, we first obtain a solution of
(\ref{HULL:eq}) for a given value of $v$. After that we get the
corresponding value of $F$ by using (\ref{US:Fv2}), i.e.,
\begin{equation}
  F=\gamma v(1+2\sum_{m=1}^\infty m^2|f_m|^2).
  \label{HULL:Fv}
\end{equation}

In the numerical approximation we solve the set of $M$ algebraic
equations (\ref{HULL:eq}) with the Newton method. The most 
time-consuming part of the computation is the calculation of $I_m$.
In order to speed up the computation we restrict ourselves to values
of the cutoff $M$ which are powers of two. Now we are able to
calculate $\{I_1,\ldots,I_M\}$ from $\{f_1,\ldots,f_M\}$ by two Fast
Fourier Transformations (FFT). The first one is an inverse FFT which
calculates the hull function in real space. Next we calculate
$\sin[\varphi+f(\varphi)]$. The second FFT yields
$\{I_1,\ldots,I_M\}$.

In order to get reliable results the value of the cutoff $M$ has to
be chosen carefully. Due to the FFT the hull function is approximated
on a lattice with lattice constant $\pi/M$. Because of
(\ref{US:dhull}) this corresponds to a time resolution $\Delta
t=\pi/(vM)$ of $x_j(t)$. The fastest time scale in the system is
given by $2\pi$ divided by the maximum phonon frequency which is 2 in
accordance with (\ref{RES:w}). Assuming that the fastest time scale
has to be resolved at least by two steps $\Delta t$ we get reliable
results only if 
\begin{equation}
  v\gtrsim \frac{2}{M}.
  \label{HULL:vmin}
\end{equation}
All results reported in this paper are obtained for $M=32$. 
Thus, $v$ has to be larger than 0.06.

For $M=1$ we get an analytic result parameterized by the modulus of $f_1$. 
We write 
\begin{equation}
  f_1=\frac{A}{2}e^{i\psi}
  \label{HULL:f1}
\end{equation}
Due to the integral representation of the Bessel functions of the
first kind we can obtain $I_1(f_1)$. Together with (\ref{HULL:eq}) we
get 
\begin{eqnarray}
 [v^2-w^2(a)]A&=&b[J_2(A)-J_0(A)]\sin\psi, \label{HULL:f1.eq1}\\
 \gamma vA&=&b[J_2(A)+J_0(A)]\cos\psi. \label{HULL:f1.eq2}
\end{eqnarray}
The elimination of $\psi$ leads to polynomial of second order in
$v^2$. Thus, we get an analytic solution parameterized by $A$:
\begin{mathletters}\label{HULL:M1}
\begin{equation}
  v(A)=\sqrt{\omega^2(a)-\frac{\tilde{\gamma}^2}{2}\pm\sqrt{
   \frac{\tilde{\gamma}^4}{4}-\omega^2(a)\tilde{\gamma}^2+
   \left(\frac{\tilde b}{A}\right)^2}},
  \label{HULL:vA}
\end{equation}
with
\begin{equation}
  \tilde{\gamma}\equiv \frac{J_0(A)-J_2(A)}{J_0(A)+J_2(A)}\,\gamma
  \label{HULL:tga}
\end{equation}
and
\begin{equation}
 \tilde{b}\equiv [J_0(A)-J_2(A)]b.
  \label{HULL:tb}
\end{equation}
Using (\ref{HULL:Fv}) we get
\begin{equation}
  F(A)=\gamma v(A)\left(1+\frac{A^2}{2}\right).
  \label{HULL:Fv1}
\end{equation}
\end{mathletters}

\section{\protect\label{LINST}Linear stability analysis of the
uniform sliding state}

In this appendix we explain our procedure to calculate numerically the
stability of the uniform sliding state. Furthermore, we derive an
approximation for $v_c$.

Plugging the Floquet-Bloch ansatz (\ref{INST:f.sol}) into the linearized
equation of motion (\ref{INST:leqm}) leads to
\begin{eqnarray}
 \lefteqn{v^2c_k''(\varphi)+(2\lambda_k+\gamma)vc_k'(\varphi)+(\lambda_k^2+
   \lambda_k\gamma)c_k(\varphi)=}\nonumber\\&&\hspace{20mm}c_k(\varphi-a)
   e^{-\frac{2\pi i}{N}k}+c_k(\varphi+a)e^{\frac{2\pi i}{N}k}-2c_k(\varphi)
   \nonumber\\&&\hspace{20mm}-b\cos[\varphi+f(\varphi)]c_k(\varphi).
  \label{LINST:c.eqm}
\end{eqnarray}
The Fourier ansatz
\begin{equation}
  c_k(\varphi)=\sum_{m=0}^\infty c_{k,m}e^{im\varphi}+\mbox{c.c.}
  \label{LINST:ckm}
\end{equation}
turns this equation into a set of infinitely many linear equations:
\begin{eqnarray}
 \lefteqn{\left[\omega^2\left(ma+\frac{2\pi k}{N}\right)+
   (\lambda_k+imv)^2+\gamma(\lambda_k+imv)\right]c_{k,m}=}\nonumber\\
   &&\hspace{20mm}-b\sum_{m'=0}^\infty
   (K_{m-m'}c_{k,m'}+K_{m+m'}c_{k,m'}^*),
  \label{LINST:ckm.eq}
\end{eqnarray}
where 
\begin{equation}
  K_m\equiv \frac{1}{2\pi}\int_0^{2\pi}\cos[\varphi+f(\varphi)]e^{-im\varphi}d\varphi.
  \label{LINST:Km}
\end{equation}

Again we have to choose a cutoff $M'$ to solve this set numerically.
In order to be consistent with the cutoff $M=32$ of the Fourier
expansion of the hull function $f$ we have chosen $M'=15$. Because of
(\ref{INST:f.sol}) the stability depends on the number $N$ of
particles. For large $N$ the eigenvalues $\lambda_k$ lead to a continuous
function $\lambda(2\pi k/N)$. Since we are mainly interested in large
systems ($N>50$) we have chosen $N=100$. It is difficult to check
numerically whether the uniform sliding state is stable or not
because of the Goldstone mode $\lambda_0=0$.  Our numerically
obtained value of $\lambda_0$ fluctuates around zero because of
unavoidable errors. Therefore our instability criterion reads: The
uniform sliding state is unstable if at least one eigenvalue is
larger than $0.05$. This value is considerably larger than the
amplitude of the numerical fluctuations of $\lambda_0$. 

We are able to obtain an analytical approximation of the largest
sliding velocity $v_c$ at which parametric resonance is just able to
destabilize the uniform sliding state. We do this for zero damping
because physical intuition tells us that $v_c$ decreases
monotonically with $\gamma$. In fact for $\gamma\ll 1$ and $b={\cal
O}(\gamma^0)$, the largest sliding velocity is the undamped one minus
a correction term of order $\gamma$. The approximation makes three
assumption: (i) Parametric resonance at $v_c$ occurs for that value
of $q$ which maximizes $v_1^P(q)$. For $0\le a\le\pi$ this implies
$q=\pi$. (ii) All Fourier coefficients of $c_k$ are zero except
$c_{k,0}$ and $c_{k,1}$. (iii) The integrals $K_m$ are calculated
only in leading order of $b$ which yields
\begin{equation}
  K_m\approx\frac{1}{2\pi}\int_0^{2\pi}\cos(\varphi)e^{-im\varphi}d\varphi
   =\frac{\delta_{1,m}+\delta_{-1,m}}{2},
  \label{LINST:Km0}
\end{equation}
where $\delta_{n,m}$ is the Kronecker symbol.
With $\gamma=0$ and these assumptions (\ref{LINST:ckm.eq}) reduces to
\begin{equation}
  \left(\begin{array}{cc}
     \omega_0^2+\lambda^2 & b/2\\ b/2 & \omega_0^2+(\lambda+iv)^2
  \end{array}\right)
  \left(\begin{array}{c}c_{k,0}\\c_{k,1}\end{array}\right)=
  \left(\begin{array}{c}0\\0\end{array}\right),
  \label{LINST:ck01.eq}
\end{equation}
where
\begin{equation}
  \omega_0\equiv\omega(k)=\omega(a+k)=\omega(a/2\pm\pi)=
   2\cos\left(\frac{a}{4}\right).
  \label{LINST:w0}
\end{equation}
A nontrivial solution of (\ref{LINST:ck01.eq}) implies a zero
determinant leading to a characteristic polynomial of second order
in $(\lambda+iv/2)^2$. Solutions with nonzero real part of $\lambda$
occur only if $(\omega_0^2-v^2/4)^2<b^2/4$. Therefore we get
(\ref{INST:vc}).


\begin{references}
\bibitem{US:rem1}
It should be noted that it is mathematically not evident that (\ref{US:DDEQ})
has a solution at all \cite{wat96}. Numerically we always found at least
one solution as long as $v$ is larger than the limit dictated by our
numerical scheme (see appendix~\ref{HULL}).
\bibitem{rem2}
The conjecture in Ref.~\onlinecite{str96} that $a_1$ and $a_2$ are
highly rational numbers is generally not true.
\end{references}
\end{document}